\documentclass[aps,prd,twocolumn,superscriptaddress,nofootinbib,floatfix]{revtex4-2}

\usepackage[T1]{fontenc}
\usepackage[utf8]{inputenc}
\usepackage{amsmath,amssymb,mathtools}
\usepackage{bm}
\usepackage{graphicx}
\usepackage[hidelinks]{hyperref}
\usepackage{orcidlink}

\begin{document}
\title{Photon-Return and Polar-Focusing Crossovers near Extremal Kerr.}

\author{Valentina González-Moya}
\email{valentina.gonzalez.m01@mail.pucv.cl}
\affiliation{Instituto de F\'\i sica, Pontificia Universidad Cat\'olica de Valpara\'\i so, Casilla 4950, Valpara\'\i so, Chile.}

\author{Ali \"Ovg\"un \orcidlink{0000-0002-9889-342X}}
\email{ali.ovgun@emu.edu.tr}
\affiliation{Physics Department, Faculty of Arts and Sciences, Eastern Mediterranean University, Famagusta, Cyprus.}

\author{Joel Saavedra\orcidlink{0000-0002-1430-3008}}
\email{joel.saavedra@pucv.cl}

\affiliation{Instituto de F\'\i sica, Pontificia Universidad Cat\'olica de Valpara\'\i so, Casilla 4950, Valpara\'\i so, Chile.}

\date{\today}
\begin{abstract}
How precisely must a photon's impact parameter be tuned for it to
complete $n$ windings around a nearly extremal Kerr black hole before
escaping, and how does the same near extremal scaling control the
focusing of neighboring rays? For equatorial prograde returns we find,
with $\kappa=\sqrt{1-a^2/M^2}$ and $X=n\kappa$,
$\epsilon_n=\sqrt{3}\,\kappa/(e^{2\pi X}-1)
+O(\kappa^2|\ln\kappa|)$. This single law connects the extremal
inverse-winding hierarchy to subextremal exponential accumulation.
Perturbing the return family quasi-equatorially, we derive a second
scaling law for the polar Jacobi phase,
$\Psi_\theta=2\ln[\sinh(\pi X)/\kappa]+C_s+o(1)$, and show that its
focusing points cross from geometric spacing in winding number to
asymptotically uniform spacing $\Delta\mathcal N\simeq(2\kappa)^{-1}$.
Numerical integrations of the exact Kerr radial dynamics and linear polar equation verify both scaling collapses. Thus near extremal
Kerr exhibits parametrically distinct winding and polar-focusing
scales, rather than a discontinuous jump between extremal and
subextremal lensing hierarchies.
\end{abstract}

\maketitle

\section{Introduction}
Black-hole photon rings arise from null trajectories that remain close
to unstable photon orbits before escaping.  For a nondegenerate
critical orbit, successive near-critical trajectories accumulate
exponentially, with the hierarchy controlled by the instability of the
underlying orbit \cite{Bozza:2002zj,Gralla:2019drh,Johnson:2019ljv}.
Strong-field Kerr lensing also couples azimuthal winding to polar
motion: quasi-equatorial treatments provide two-dimensional lens maps,
magnifications, and caustics \cite{Bozza:2002af}, while the exact Kerr
geodesic problem admits a fully separated analytic description
\cite{Gralla:2019ceu}.

Rapid rotation produces a singular limit.  As $a/M\to1$, the prograde
equatorial photon orbit approaches the degenerate horizon and its
coordinate-time instability per revolution becomes parametrically
small.  Near-horizon Kerr geodesics consequently develop their own
scaling structure \cite{Compere:2020eat,Gates:2020els}.
For the extremal prograde orbit the usual logarithmic strong-deflection
expansion is replaced by a stronger power-law contribution; the
resulting equatorial hierarchy is algebraic rather than exponential
\cite{Barlow:2017uty,Feleppa:2026emm}.  Recent work has also constructed quasi-equatorial magnifications
and caustics directly at extremality~\cite{Feleppa:2026emm}, revealing a
qualitatively different higher-order caustic structure from the
standard nonextremal strong-deflection regime. In that endpoint problem
the caustic displacement is exponential in the polar index, whereas
the standard nonextremal quasi-equatorial hierarchy is linear in that
index \cite{Bozza:2002af,Feleppa:2026emm}.  These endpoint
descriptions, however, do not by themselves determine the controlled
near extremal interpolation when the azimuthal winding diverges
simultaneously with the approach to extremality.

This work identifies and resolves a controlled nonuniform
near extremal limit for a fixed exterior return radius.  We introduce
\begin{equation}
 \kappa=\sqrt{1-\frac{a^2}{M^2}},
 \qquad
 X=n\kappa,
 \label{eq:scaling-variable}
\end{equation}
where $n$ counts complete azimuthal windings.  Starting from the exact
Kerr equations, we show that the impact-parameter offset from the
prograde critical orbit obeys
\begin{equation}
 \epsilon_n
 =\frac{\sqrt{3}\,\kappa}{e^{2\pi X}-1}
 +O\!\left(\kappa^2|\ln\kappa|\right),
 \label{eq:intro-scaling}
\end{equation}
for fixed positive finite $X$.  The same double scaling controls a
second quantity: the linear polar phase accumulated by neighboring
quasi-equatorial rays.  Its focusing hierarchy possesses a distinct
crossover order $m_{\rm cross}^{(\theta)}\sim(2/\pi)\ln\kappa^{-1}$,
whereas the azimuthal return crossover occurs at
$n_{\rm cross}\sim\kappa^{-1}$.  Exact Kerr integrations verify both
scaling laws.  The result therefore identifies two parametrically
different geometrical scales in near extremal lensing.

\section{Near extremal return hierarchy}
\label{sec:return-hierarchy}

We consider equatorial null geodesics of a Kerr black hole
\cite{Carter:1968rr,Gralla:2019ceu}. Setting the conserved energy to
unity, the radial potential and azimuthal evolution are
\begin{align}
 R(r)&=(r^2+a^2-a\xi)^2-\Delta(\xi-a)^2,
 \nonumber\\[-2pt]
 \Delta&=r^2-2Mr+a^2,
 \label{eq:radial-potential}\\
 \frac{d\phi}{dr}
 &=\frac{\xi-a}{\pm\sqrt{R(r)}}
 \nonumber\\[-2pt]
 &\quad+\frac{a(r^2+a^2-a\xi)}
 {\pm\Delta\sqrt{R(r)}}.
 \label{eq:dphidr}
\end{align}
where $\xi=L_z/E$. For an exterior prograde trajectory with turning
point $r_0$, $R(r_0)=0$ gives
\begin{equation}
 \xi=a+\frac{r_0^2}{a+\sqrt{\Delta(r_0)}}.
 \label{eq:xi-r0}
\end{equation}
The $n$th returning trajectory is then defined exactly by
\begin{equation}
 2\int_{r_0}^{r_s}
 \left|\frac{d\phi}{dr}\right|dr=2\pi n.
 \label{eq:return-condition}
\end{equation}

To resolve the near extremal region we use
\begin{align}
 a&=M\sqrt{1-\kappa^2},
 \qquad r=M(1+\kappa y),
 \nonumber\\
 r_0&=M(1+\kappa y_0).
 \label{eq:near-scaling}
\end{align}
and write $s_0=\sqrt{y_0^2-1}$. \paragraph{Exact turning-point expansion.}
\begin{equation}
 \Delta_0=M^2\kappa^2s_0^2,
\end{equation}
and Eq.~\eqref{eq:xi-r0} becomes
\begin{equation}
 \frac{\xi}{M}
 =
 \sqrt{1-\kappa^2}
 +
 \frac{(1+\kappa y_0)^2}
 {\sqrt{1-\kappa^2}+\kappa s_0}.
 \label{eq:supp-xi-exact}
\end{equation}
Expanding at fixed $y_0$ gives
\begin{align}
 \frac{\xi}{M}
 &=
 2+\kappa(2y_0-s_0)
 \nonumber\\
 &\quad
 +\kappa^2
 \left(2y_0^2-2y_0s_0-1\right)
 +O(\kappa^3).
 \label{eq:supp-xi-expansion}
\end{align}

The turning-point relation and the
critical prograde orbit give
\begin{align}
 \frac{\xi}{M}
 &=2+\kappa(2y_0-s_0)+O(\kappa^2),
 \nonumber\\
 \frac{r_{\rm ph}}{M}
 &=1+\frac{2}{\sqrt{3}}\kappa+O(\kappa^2),
 \qquad
 \frac{\xi_c}{M}
 =2+\sqrt{3}\,\kappa+O(\kappa^2).
 \label{eq:critical-expansions}
\end{align}
Hence, defining $\epsilon=(\xi-\xi_c)/\xi_c$,
\begin{equation}
 \frac{\epsilon}{\kappa}
 =
 h(y_0)+O(\kappa),
 \qquad
 h(y_0)=
 \frac{2y_0-\sqrt{y_0^2-1}-\sqrt{3}}{2}.
 \label{eq:h-def}
\end{equation}

The same scaling produces a nontrivial near-horizon limit of the
radial dynamics. Substituting Eq.~\eqref{eq:near-scaling} into the
exact potential gives
\begin{equation}
 R(r)
 =
 M^4\kappa^2
 (y-y_0)
 \left(3y-5y_0+4s_0\right)
 +O(\kappa^3).
 \label{eq:R-scaling}
\end{equation}
The factorization of the leading term is crucial: it retains both the
turning point $y_0$ and the critical photon orbit at finite locations
in the scaled coordinate. Combining this result with the exact
azimuthal equation yields
\begin{equation}
 d\phi
 =
 \frac{dy}{\kappa}
 \frac{2y-2y_0+s_0}
 {(y^2-1)
 \sqrt{(y-y_0)(3y-5y_0+4s_0)}}
 +O(1)\,dy .
 \label{eq:dphi-scaling}
\end{equation}
The $1/\kappa$ enhancement is the origin of the nonuniform
large-winding limit. In particular, a finite scaled azimuthal
integral can balance the return condition only when the winding
number grows as $n=O(\kappa^{-1})$.

At the same order, the complete return angle becomes
\begin{equation}
 \Delta\phi_{\rm ret}
 =
 \frac{2}{\kappa}J(y_0)
 +O(|\ln\kappa|),
 \label{eq:return-scaling}
\end{equation}
where, for the exterior prograde branch
$y_0>2/\sqrt{3}$,
\begin{equation}
 J(y_0)
 =
 \int_{y_0}^{\infty}
 \frac{
 2y-2y_0+s_0
 }{
 (y^2-1)
 \sqrt{(y-y_0)(3y-5y_0+4s_0)}
 }\,dy .
 \label{eq:J-def}
\end{equation}

For a source radius $r_s/M>1$ held fixed as $\kappa\to0$, the
scaled upper limit tends to infinity. Matching the near-horizon
solution to the exterior region fixes the first subleading
contribution,
\begin{equation}
 \Delta\phi_{\rm ret}
 =
 \frac{2}{\kappa}J(y_0)
 -\frac{4}{3\sqrt{3}}\ln\kappa
 +O(1).
 \label{eq:return-matched}
\end{equation}
The explicit matched-asymptotic derivation is presented below. Imposing the exact
return condition then gives
\begin{equation}
 J(y_0)
 =
 \pi X+
 \frac{2\kappa}{3\sqrt{3}}\ln\kappa
 +O(\kappa),
 \qquad
 X=n\kappa .
 \label{eq:J-matched}
\end{equation}
Thus, at leading order in the double-scaling limit,
\begin{equation}
 J[y_0(X)]=\pi X,
 \qquad
 H(X)=h[y_0(X)],
 \label{eq:scaling-function}
\end{equation}
The integral is elementary on this branch.  Writing
$p=2y_0-s_0$, one finds
\begin{equation}
 J(y_0)=\operatorname{artanh}\!\left(\frac{\sqrt{3}}{p}\right),
 \qquad
 p=\sqrt{3}\,\coth(\pi X).
 \label{eq:J-closed}
\end{equation}
\paragraph{Evaluation of the radial integral.}
  Define
\begin{equation}
 p=2y_0-s_0,
 \qquad
 D(y)=3y^2-4py+p^2+1.
 \label{eq:supp-pD}
\end{equation}
Using
\begin{equation}
 \frac{d}{dy}\operatorname{artanh}
 \!\left[\frac{\sqrt{D(y)}}{py-2}\right]
 =\frac{2y-p}{(y^2-1)\sqrt{D(y)}},
 \label{eq:supp-Jprimitive}
\end{equation}
This primitive yields the closed expression in Eq.~\eqref{eq:J-closed}.

Since $h(y_0)=(p-\sqrt{3})/2$, the crossover function is explicit,
\begin{equation}
H(X)=\frac{\sqrt{3}}{e^{2\pi X}-1}.
 \label{eq:H-closed}
\end{equation}
Equation~\eqref{eq:H-closed}, together with the matched correction in
Eq.~\eqref{eq:J-matched}, yields Eq.~\eqref{eq:intro-scaling}.  The
quoted remainder is controlled at fixed positive finite $X$; no
uniformity is assumed as $X\to0$ or $X\to\infty$.

\subsection{Matched asymptotic expansion at fixed exterior radius}
We now determine the first subleading contribution to the return
angle for a source radius $r_s/M>1$ held fixed as
$\kappa\to0$. The near-horizon expansion derived above is not uniform
up to the source, since
\begin{equation}
 y_s=\frac{r_s/M-1}{\kappa}=O(\kappa^{-1}).
\end{equation}
A matched asymptotic expansion is therefore required. We introduce
an intermediate radius
\begin{equation}
r_m=M(1+\kappa Y),
\qquad
Y=\kappa^{-\alpha},
\qquad
\frac{1}{2}<\alpha<1,
\label{eq:matching-point}
\end{equation}
or equivalently $x_m=\kappa Y$, where $x=r/M-1$.
This choice lies in the overlap region
$1\ll Y\ll\kappa^{-1}$ and ensures that
$1/(\kappa Y^2)$, $1/Y$, and $\kappa Y$ all vanish as
$\kappa\to0$. The return integral is split into near and far
contributions at $r_m$.

\emph{Near region.}
The complete expression for $F_1$ is algebraically lengthy and is not
needed for the matching. Its large-$y$ behavior in the overlap region
is
\begin{align}
 F_0(y;y_0)&=\frac{2}{\sqrt{3}\,y^2}+O(y^{-3}),\nonumber\\
 F_1(y;y_0)&=\frac{2}{3\sqrt{3}\,y}+O(y^{-2}).
 \label{eq:supp-inner-overlap}
\end{align}
Importantly, the coefficient of the $1/y$ term is independent of the
turning point $y_0$. Integrating from $y_0$ to the matching point
$Y$ gives, for one radial branch,
\begin{align}
I_{\rm near}
&=
\frac{1}{\kappa}
\int_{y_0}^{Y}F_0(y;y_0)\,dy
+
\int_{y_0}^{Y}F_1(y;y_0)\,dy
+O(1)
\nonumber\\
&=
\frac{J(y_0)}{\kappa}
-\frac{2}{\sqrt{3}\,\kappa Y}
+\frac{2}{3\sqrt{3}}\ln Y
+C_{\rm near}(y_0)
\nonumber\\
&\quad
+O\left(\frac{1}{\kappa Y^2}\right)
+O\left(\frac{1}{Y}\right)
+O(\kappa Y).
\label{eq:supp-Inear}
\end{align}
where $C_{\rm near}(y_0)$ denotes terms finite in the matching limit.
The first correction follows from the large-$Y$ tail
\begin{equation}
 \int_Y^\infty F_0(y;y_0)\,dy
 =
 \frac{2}{\sqrt{3}\,Y}+O(Y^{-2}).
 \label{eq:supp-F0-tail}
\end{equation}

\emph{Far region.}
In the exterior region it is convenient to keep
\begin{equation}
 x=\frac{r}{M}-1
\end{equation}
fixed as $\kappa\to0$. At leading order, the impact parameter tends
to its extremal critical value $\xi_c=2M$, and the azimuthal
integrand for one radial branch reduces to
\begin{equation}
 G_0(x)
 =
 \frac{2\sqrt{x+1}}
 {x^2\sqrt{x+3}}.
 \label{eq:supp-G0}
\end{equation}
Its small-$x$ expansion is
\begin{equation}
 G_0(x)
 =
 \frac{2}{\sqrt{3}\,x^2}
 +\frac{2}{3\sqrt{3}}\frac{1}{x}
 -\frac{1}{3\sqrt{3}}
 +O(x).
 \label{eq:supp-G0-overlap}
\end{equation}
The first two terms reproduce precisely the overlap structure of the
near-region expansion after using $x=\kappa y$.

For one branch, the far contribution is
\begin{equation}
 I_{\rm far}
=
\frac{2}{\sqrt{3}\,\kappa Y}
-\frac{2}{3\sqrt{3}}\ln(\kappa Y)
+C_{\rm far}(r_s)
+o(1),
 \label{eq:supp-Ifar}
\end{equation}
where corrections to the leading far integrand contribute only to
the finite remainder at the order required here. 
The finite quantity $C_{\rm far}(r_s)$ depends on the fixed source
radius but does not affect the logarithmically enhanced correction.

\emph{Matching.}
Adding Eqs.~\eqref{eq:supp-Inear} and
\eqref{eq:supp-Ifar}, the inverse powers of the arbitrary matching
scale cancel,
\begin{equation}
 -\frac{2}{\sqrt{3}\,\kappa Y}
 +
 \frac{2}{\sqrt{3}\,\kappa Y}
 =0,
\end{equation}
and so do the logarithms of $Y$,
\begin{equation}
 \frac{2}{3\sqrt{3}}\ln Y
 -
 \frac{2}{3\sqrt{3}}\ln(\kappa Y)
 =
 -\frac{2}{3\sqrt{3}}\ln\kappa.
\end{equation}
Thus the azimuthal angle accumulated along one radial branch is
\begin{equation}
 I
 =
 \frac{J(y_0)}{\kappa}
 -\frac{2}{3\sqrt{3}}\ln\kappa
 +C(y_0,r_s)
 +o(1),
 \label{eq:supp-one-branch-matched}
\end{equation}
where $C=C_{\rm near}+C_{\rm far}$ is finite as
$\kappa\to0$. Doubling this one-leg result reproduces Eq.~\eqref{eq:return-matched}.

The source-radius dependence enters only through the finite matching
constant $C(y_0,r_s)$ and therefore does not modify $H(X)$ at leading
order for fixed $r_s/M>1$. This conclusion changes if the source is
itself scaled into the near-horizon region. If
\begin{equation}
 r_s-M=O(\kappa M),
\end{equation}
then $y_s=O(1)$ rather than $y_s\to\infty$, and the upper limit of the
near-horizon integral remains finite.

\section{Algebraic to exponential crossover}
\label{sec:crossover}

The scaling function has two asymptotic regimes. For $X\ll1$,
Eq.~\eqref{eq:scaling-function} selects $y_0\gg1$, where
\begin{equation}
 J(y_0)\sim\frac{\sqrt{3}}{y_0},
 \qquad
 h(y_0)\sim\frac{y_0}{2}.
 \label{eq:smallX}
\end{equation}
It follows that
\begin{equation}
 H(X)\sim\frac{\sqrt{3}}{2\pi X},
 \qquad
 \epsilon_n\sim\frac{\sqrt{3}}{2\pi n},
 \qquad X\ll1.
 \label{eq:algebraic}
\end{equation}
Within the large-winding double-scaling regime, this corresponds to
$1\ll n\ll\kappa^{-1}$.
This reproduces the algebraic inverse-winding behavior associated
with the extremal prograde strong-deflection limit
\cite{Feleppa:2026emm}. Thus, at sufficiently small
$n\kappa$, a strictly subextremal trajectory retains the leading
extremal hierarchy.

For $X\gg1$, the turning point approaches the scaled photon orbit.
Writing $y_0=2/\sqrt{3}+\eta$ with $\eta\ll1$, one finds
\begin{align}
 J(y_0)
 &=-\ln\eta+\ln\left(\frac{2}{\sqrt{3}}\right)
 +O(\eta\ln\eta),
 \nonumber\\
 h(y_0)&=\frac{3\sqrt{3}}{4}\eta^2+O(\eta^3).
 \label{eq:largeX-expansion}
\end{align}
Equation~\eqref{eq:scaling-function} then gives
\begin{equation}
 H(X)\sim\sqrt{3}\,e^{-2\pi X},
 \qquad
 \epsilon_n\sim\sqrt{3}\,\kappa e^{-2\pi n\kappa},
 \qquad X\gg1.
 \label{eq:exponential}
\end{equation}

The algebraic and exponential hierarchies are therefore the two
asymptotic sectors of a single scaling function. Their crossover
occurs at $X=O(1)$, corresponding to
\begin{equation}
 n_{\rm cross}\sim\kappa^{-1}
 =
 \left(1-\frac{a^2}{M^2}\right)^{-1/2}.
 \label{eq:ncross}
\end{equation}
The extremal limit is therefore nonuniform in the winding number.
Within the large-winding regime, the route
$n\to\infty$, $\kappa\to0$ with $n\kappa\to0$ selects the algebraic
sector, whereas increasing $n$ at any fixed $\kappa>0$ ultimately
selects exponential accumulation.

\begin{figure}[t]
 \centering
 \includegraphics[width=\columnwidth]
 {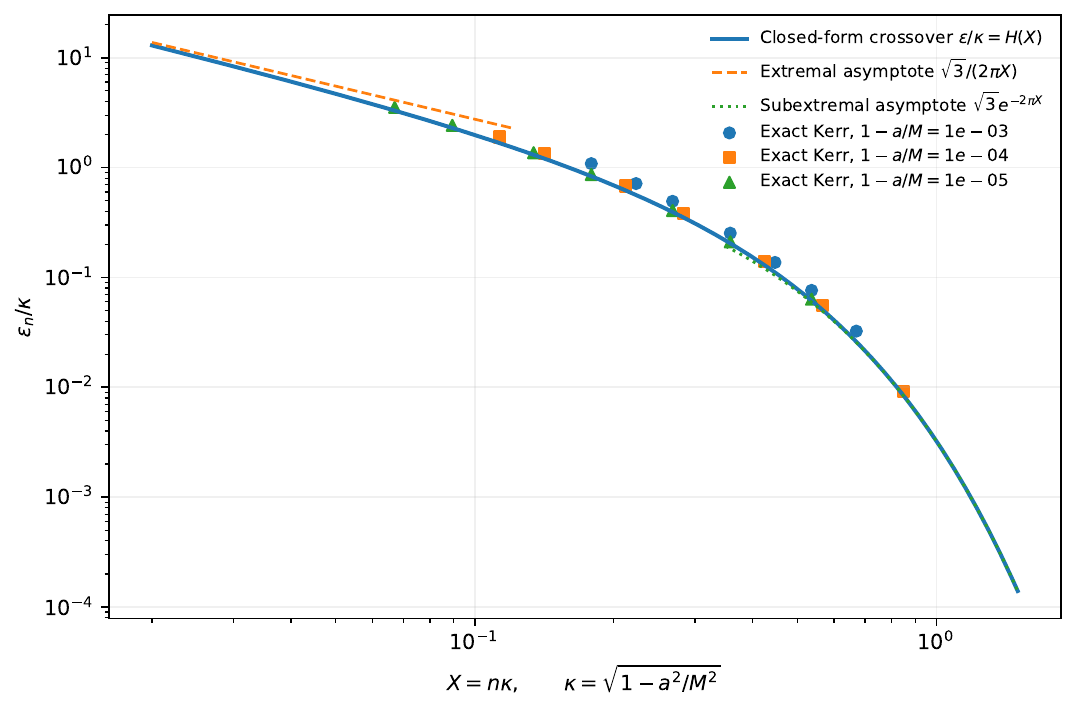}
 \caption{
 Scaling collapse of equatorial prograde Kerr return geodesics at
 $r_s=100M$. Exact geodesic sequences approach the parameter-free
 function $H(X)$ as $\kappa\to0$. The dashed curves show the
 algebraic and exponential asymptotes in
 Eqs.~\eqref{eq:algebraic} and \eqref{eq:exponential}.
 }
 \label{fig:crossover}
\end{figure}

We test this prediction by solving the exact return condition
\eqref{eq:return-condition}, without using the near extremal
expansion. Figure~\ref{fig:crossover} shows $\epsilon_n/\kappa$
against $X=n\kappa$ for several near extremal spins at fixed
$r_s=100M$. As $\kappa$ decreases, the exact Kerr sequences collapse
onto the parameter-free scaling function defined by
Eq.~\eqref{eq:scaling-function}, while approaching the algebraic and
exponential limits in Eqs.~\eqref{eq:algebraic} and
\eqref{eq:exponential}.

\subsection{Numerical validation and the logarithmic correction}

\medskip
The exact return condition is evaluated without using the
near extremal expansion. To regularize the integrable square-root
singularity at the turning point, we use
\begin{equation}
r=r_0+(r_s-r_0)z^2,
\qquad 0\leq z\leq1.
\end{equation}
For numerical stability near $r=r_0$, defining $q=\xi-a$ allows the
radial potential to be factorized exactly as
\begin{equation}
R(r)=
r(r-r_0)
\left[
r^2+r_0r+r_0^2-(q^2+2aq)
\right].
\end{equation}
For each $(n,\kappa)$, we solve the exact condition
$\Delta\phi_{\rm ret}=2\pi n$ for $r_0$ and compute the
corresponding impact-parameter offset directly from the exact Kerr
expressions.

To test the subleading matched expansion, we keep $X=n\kappa$ fixed
while increasing $n$, so that
\begin{equation}
\kappa=\frac{X}{n}\longrightarrow0.
\end{equation}
For each member of this sequence, the exact return condition is
solved as described above and the corresponding scaled turning point
is defined by
\begin{equation}
 y_0^{\rm exact}
 =
 \frac{r_{0,n}/M-1}{\kappa}.
 \label{eq:supp-y0-exact}
\end{equation}
Equation~\eqref{eq:J-matched} predicts
\begin{equation}
 J(y_0^{\rm exact})-\pi X
 =
 \frac{2\kappa}{3\sqrt{3}}\ln\kappa
 +O(\kappa).
 \label{eq:supp-J-correction}
\end{equation}
We therefore define the effective logarithmic coefficient
\begin{equation}
 C_{\rm eff}(\kappa;X)
 =
 \frac{
 J(y_0^{\rm exact})-\pi X
 }{
 \kappa\ln\kappa
 }.
 \label{eq:supp-Ceff}
\end{equation}
For every fixed finite $X$ in the double-scaling regime,
the matched expansion predicts
\begin{equation}
 \lim_{\kappa\to0}C_{\rm eff}(\kappa;X)
 =
 \frac{2}{3\sqrt{3}}
 \simeq0.384900.
 \label{eq:supp-Ceff-limit}
\end{equation}
Because the unshown term in
Eq.~\eqref{eq:supp-J-correction} is $O(\kappa)$, the convergence of
$C_{\rm eff}$ is only logarithmic,
\begin{equation}
 C_{\rm eff}
 =
 \frac{2}{3\sqrt{3}}
 +O\left(\frac{1}{|\ln\kappa|}\right).
 \label{eq:supp-Ceff-convergence}
\end{equation}
This slow convergence is expected and provides a more stringent
test of the matched expansion than the leading scaling collapse.
The same correction can be propagated to the critical
impact-parameter offset. Let $y_*(X)$ denote the leading-order
solution
\begin{equation}
 J(y_*)=\pi X.
 \label{eq:supp-ystar}
\end{equation}
Expanding Eq.~\eqref{eq:J-matched} about $y_*$ gives
\begin{equation}
 y_0-y_*
 =
 \frac{2\kappa}{3\sqrt{3}}
 \frac{\ln\kappa}{J'(y_*)}
 +O(\kappa).
 \label{eq:supp-y-correction}
\end{equation}
Using $\epsilon=\kappa h(y_0)+O(\kappa^2)$ then yields
\begin{equation}
 \epsilon_n
 =
 \kappa H(X)
 +
 \kappa^2\ln\kappa\,K(X)
 +O(\kappa^2),
 \label{eq:supp-epsilon-subleading}
\end{equation}
where
\begin{align}
 K(X)
 &=\frac{2}{3\sqrt{3}}
 \frac{h'[y_*(X)]}{J'[y_*(X)]}
 =\frac{2}{3\sqrt3\,\pi}H'(X)
 \nonumber\\
 &=-\frac{1}{3\sinh^2(\pi X)}.
 \label{eq:supp-K}
\end{align}
Thus the first logarithmically enhanced correction can also be written
explicitly as
\begin{equation}
 \epsilon_n
 =\frac{\sqrt3\,\kappa}{e^{2\pi X}-1}
 -\frac{\kappa^2\ln\kappa}{3\sinh^2(\pi X)}
 +O(\kappa^2).
 \label{eq:supp-epsilon-explicit-subleading}
\end{equation}
Thus the normalized exact-geodesic residual
\begin{equation}
 {\cal R}_{\epsilon}(\kappa;X)
 =
 \frac{
 \epsilon_n^{\rm exact}-\kappa H(X)
 }{
 \kappa^2\ln\kappa
 }
 \label{eq:supp-residual}
\end{equation}
must approach $K(X)$ as $\kappa\to0$ at fixed $X$.
\begin{figure}[t]
\centering
\includegraphics[width=0.95\columnwidth]{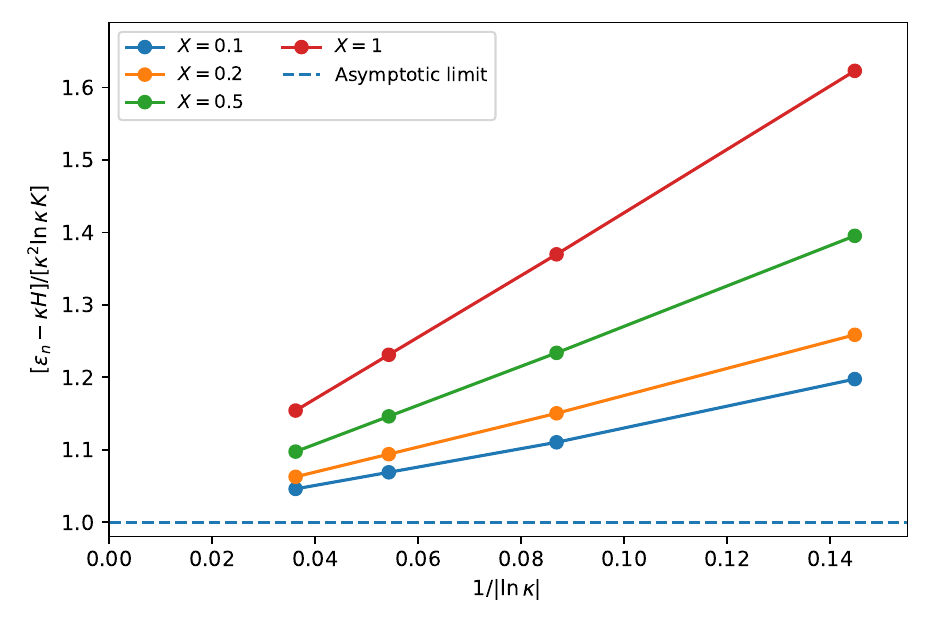}
\caption{
Numerical test of the logarithmically enhanced correction at fixed
$X$. The normalized residual
${\cal R}_{\epsilon}/K(X)$ is shown as a function of
$1/|\ln\kappa|$ for representative values of the double-scaling
variable. The exact Kerr return sequences approach the matched
asymptotic prediction ${\cal R}_{\epsilon}/K(X)=1$ as
$\kappa\to0$. The slow convergence is consistent with the
$O(1/|\ln\kappa|)$ correction expected from
Eq.~\eqref{eq:supp-epsilon-subleading}.
}
\label{fig:supp-log}
\end{figure}
The exact numerical sequences exhibit this behavior, providing an
independent check of both the scaling function and the coefficient
of the logarithmically enhanced correction.

\section{Polar phase and focusing crossover}
\label{sec:polar-focusing}

The return hierarchy is equatorial, but its stability to polar
perturbations supplies a second geometrical observable.  For a general
null Kerr geodesic, the separated polar potential is
\begin{equation}
 \Theta(\theta)=\eta+a^2\cos^2\theta-\xi^2\cot^2\theta,
 \qquad \eta=Q/E^2,
 \label{eq:theta-potential}
\end{equation}
where $Q$ is the Carter constant.  Using Mino parameter $\gamma$,
defined by $d\gamma=d\lambda_{\rm aff}/\Sigma$, and perturbing an
equatorial trajectory by
$\vartheta=\theta-\pi/2$, Eq.~\eqref{eq:theta-potential} gives
\begin{equation}
 \frac{d^2\vartheta}{d\gamma^2}
 +\omega_\theta^2\vartheta=O(\vartheta^3),
 \qquad
 \omega_\theta=\sqrt{\xi^2-a^2}.
 \label{eq:polar-oscillator}
\end{equation}
Thus the exact linear polar phase accumulated along an equatorial
return is
\begin{equation}
 \Psi_\theta
 =2\omega_\theta\int_{r_0}^{r_s}\frac{dr}{\sqrt{R(r)}}.
 \label{eq:polar-phase-exact}
\end{equation}
This is a Jacobi phase about the equatorial geodesic; it should not be
confused with a full two-dimensional lens-map magnification.

\subsection{Near-far matching of the polar phase}
For the matching calculation we set $M=1$ temporarily.  In the inner
region, Eq.~\eqref{eq:R-scaling} gives the one-leg Mino integral
\begin{align}
 I_{\rm in}
 &=\int_{y_0}^{Y}
 \frac{dy}{\sqrt{(y-y_0)(3y-5y_0+4s_0)}}
 \nonumber\\
 &=\frac{2}{\sqrt3}\operatorname{arsinh}
 \sqrt{\frac{3(Y-y_0)}{2(2s_0-y_0)}}
 \nonumber\\
 &=\frac{1}{\sqrt3}
 \ln\!\left[\frac{6Y}{2s_0-y_0}\right]+o(1),
 \label{eq:supp-polar-inner}
\end{align}
where $1\ll Y\ll\kappa^{-1}$.  In the far region, the leading
extremal values are $a=1$ and $\xi=2$, so with $x=r-1$,
\begin{equation}
 R=x^2(x+1)(x+3).
\end{equation}
The required primitive is elementary,
\begin{equation}
 \int\frac{dx}{x\sqrt{(x+1)(x+3)}}
 =-\frac{2}{\sqrt3}\operatorname{artanh}
 \sqrt{\frac{x+3}{3(x+1)}}.
 \label{eq:supp-polar-far-primitive}
\end{equation}
For $x_m=\kappa Y$ and $\rho_s=r_s/M$, this yields
\begin{equation}
 I_{\rm far}
 =\frac{1}{\sqrt3}\ln\!\left(\frac{6}{\kappa Y}\right)
 -\frac{2}{\sqrt3}\operatorname{artanh}
 \sqrt{\frac{\rho_s+2}{3\rho_s}}+o(1).
 \label{eq:supp-polar-far}
\end{equation}
The arbitrary matching scale cancels between
Eqs.~\eqref{eq:supp-polar-inner} and \eqref{eq:supp-polar-far}, giving
for one radial leg
\begin{equation}
 I_\gamma
 =\frac{1}{\sqrt3}
 \ln\!\left[\frac{36}{\kappa(2s_0-y_0)}\right]
 -\frac{2}{\sqrt3}\operatorname{artanh}
 \sqrt{\frac{\rho_s+2}{3\rho_s}}+o(1).
 \label{eq:supp-polar-matched}
\end{equation}
Since $\omega_\theta=\sqrt3+O(\kappa)$, the $O(\kappa)$ correction to
$\omega_\theta$ multiplies only a logarithmic divergence and therefore
vanishes in the fixed-$X$ limit.

The return equation provides the final simplification.  Define
$p=2y_0-s_0$.  The leading azimuthal integral obeys
\begin{equation}
 J(y_0)=\operatorname{artanh}\!\left(\frac{\sqrt3}{p}\right)
 =\pi X,
\end{equation}
so that
\begin{equation}
 p=\sqrt3\coth(\pi X).
\end{equation}
Using the exact identity
\begin{equation}
 (2y_0-s_0)^2-(2s_0-y_0)^2=3,
\end{equation}
one obtains
\begin{equation}
 2s_0-y_0=\sqrt3\,\operatorname{csch}(\pi X).
 \label{eq:supp-q-X}
\end{equation}

The integral in Eq.~\eqref{eq:polar-phase-exact} has a different
near extremal scaling from the azimuthal angle.  Matching the
near-horizon integral to the fixed exterior endpoint gives, for
$\rho_s=r_s/M>1$ and fixed positive finite $X$,
\begin{align}
 \Psi_\theta
 &=2\ln\!\left[\frac{\sinh(\pi X)}{\kappa}\right]
 +C_s+o(1),
 \label{eq:polar-phase-scaling}\\
 C_s
 &=2\ln(12\sqrt{3})
 -4\operatorname{artanh}
 \sqrt{\frac{\rho_s+2}{3\rho_s}}.
 \label{eq:Cs}
\end{align}
The matched radial-integral derivation is presented below.  Combining
Eqs.~\eqref{eq:H-closed} and \eqref{eq:polar-phase-scaling} eliminates
$X$ and directly couples the radial tuning to polar focusing,
\begin{equation}
 \Psi_\theta
 =C_s+2\ln\!\left[
 \frac{\sqrt{3}}
 {2\sqrt{\epsilon_n(\epsilon_n+\sqrt{3}\kappa)}}
 \right]+o(1).
 \label{eq:phase-epsilon}
\end{equation}
Here $\epsilon_n$ in the logarithm is understood at leading
double-scaling order.

The linear transfer of a small polar displacement is
\begin{equation}
 \vartheta_f
 =\vartheta_i\cos\Psi_\theta
 +\frac{u_i}{\omega_\theta}\sin\Psi_\theta,
 \qquad
 u_i=\left.\frac{d\vartheta}{d\gamma}\right|_i .
 \label{eq:polar-transfer}
\end{equation}
For an equatorial launch, $\vartheta_i=0$, neighboring rays refocus in
the polar direction when $\Psi_\theta=m\pi$.  To locate these
conjugate points independently of the integer-return condition, define
the continuous unwrapped azimuthal winding
$\mathcal N=\Delta\phi/(2\pi)$.  Equations~\eqref{eq:polar-phase-scaling}
and $\Psi_\theta=m\pi$ give the focusing crossover
\begin{equation}
Y_m=\operatorname{arsinh} Z_m+o(1).
 \label{eq:focusing-crossover}
\end{equation}
Here
\begin{equation}
 Y_m=\pi\kappa\mathcal N_m,
 \qquad
 Z_m=\kappa e^{(m\pi-C_s)/2}.
 \label{eq:focusing-variables}
\end{equation}
For $Z_m\ll1$, successive focusing locations are geometrically spaced,
$\mathcal N_{m+1}/\mathcal N_m\simeq e^{\pi/2}$, reproducing the
extremal Kerr polar exponent $\hat a=2$ found in the strong-deflection
caustic analysis of Ref.~\cite{Feleppa:2026emm}.  For
$Z_m\gg1$, they become uniformly spaced,
\begin{equation}
 \mathcal N_{m+1}-\mathcal N_m\simeq\frac{1}{2\kappa}.
 \label{eq:focusing-spacing}
\end{equation}
The focusing-order crossover therefore occurs at
\begin{equation}
 m_{\rm cross}^{(\theta)}
 =\frac{2}{\pi}\ln\frac{1}{\kappa}
 +\frac{C_s}{\pi}+O(1),
 \label{eq:m-cross}
\end{equation}
which grows only logarithmically as extremality is approached, in
contrast with $n_{\rm cross}\sim\kappa^{-1}$.

\begin{figure}[t]
 \centering
 \includegraphics[width=\columnwidth]{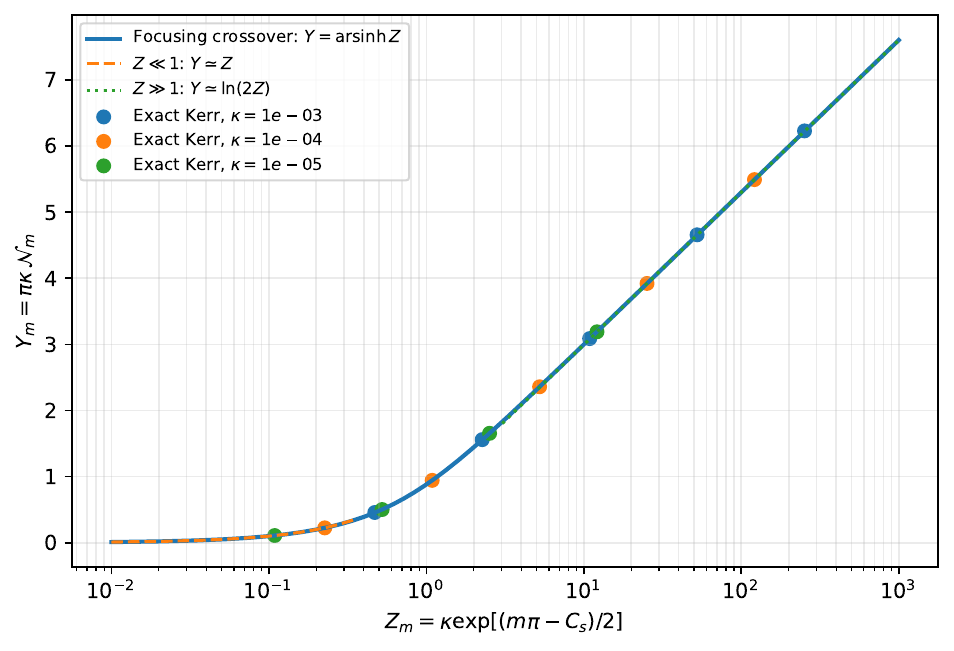}
 \caption{Polar-focusing crossover at $r_s=100M$.  The points use exact
 Kerr radial geodesics and the exact linear polar phase
 Eq.~\eqref{eq:polar-phase-exact}; only the final scaling curve uses the
 near extremal expansion.  The solid curve is
 $Y_m=\operatorname{arsinh}Z_m$, with its small- and large-$Z_m$
 asymptotes.}
 \label{fig:polar-focusing}
\end{figure}

Figure~\ref{fig:polar-focusing} tests Eq.~\eqref{eq:focusing-crossover}
without using the near extremal expansion in the radial dynamics.  We
vary the exact Kerr turning point continuously, evaluate
Eq.~\eqref{eq:polar-phase-exact}, solve $\Psi_\theta=m\pi$, and then
compute the exact unwrapped winding $\mathcal N$.  The rescaled points
collapse onto the parameter-free $\operatorname{arsinh}$ law.

The exact checks use the same turning-point regularization as the
return calculation.  For a given exact equatorial orbit we compute
$\Psi_\theta$ directly from Eq.~\eqref{eq:polar-phase-exact}, without
using the near extremal expansion.  At $r_s=100M$ and fixed $X=0.5$,
we obtain
\begin{center}
\begin{tabular}{cccc}
\hline\hline
$\kappa$ & $n$ & $\Psi_\theta^{\rm exact}$ &
$\Psi_\theta^{\rm asym}$ \\
\hline
$10^{-3}$ & $500$   & $18.874997$ & $18.882325$ \\
$10^{-4}$ & $5000$  & $23.486569$ & $23.487495$ \\
$10^{-5}$ & $50000$ & $28.092553$ & $28.092665$ \\
\hline\hline
\end{tabular}
\end{center}
The discrepancy decreases by approximately one order of magnitude for
each decade in $\kappa$ over this sequence.

For the focusing test, the turning point is varied continuously.  For
each $\kappa$ and integer $m$, we solve the exact linear condition
$\Psi_\theta=m\pi$ and evaluate the exact unwrapped winding
$\mathcal N$.  The points plotted in Fig.~\ref{fig:polar-focusing} use
$\kappa=10^{-3},10^{-4},10^{-5}$.  Across the displayed set, the
largest relative difference between
$Y_m^{\rm exact}=\pi\kappa\mathcal N_m$ and
$\operatorname{arsinh}Z_m$ is $6.1\times10^{-3}$; it decreases toward
zero as the near extremal scaling limit is approached.  As a
high-precision spot check, for $\kappa=10^{-4}$ and $m=8$ the exact
root gives
\begin{equation}
 \mathcal N_8=7507.93429426645\ldots,
\end{equation}
while the residual in the focusing condition is below
$5\times10^{-13}$ radians at 55-digit working precision.
The double-scaling prediction applies as $\kappa\to0$ with
$X=\kappa N$ finite and positive. A sequence with fixed polar
order $m$ does not, in general, realize this limit, since then
$X\to0$, where the expansion leading to Eq.~\eqref{eq:polar-phase-scaling} is not uniform.
For the numerical test we therefore choose polar orders around
and above the crossover scale
$m_{\rm cross}^{(\theta)}\sim(2/\pi)\ln(1/\kappa)$.
Figure~\ref{fig:polar-focusing} compares the resulting exact Kerr
sequences with the closed prediction $Y=\operatorname{arsinh}Z$.
The agreement systematically improves toward extremality and is
already close throughout the crossover region $Z_m=O(1)$.

\section{Dynamical origin of the winding crossover}
\label{sec:dynamical-origin}

The variable $X$ has a direct interpretation in terms of the
coordinate-time instability of the prograde circular photon orbit
\cite{Cardoso:2008bp,Johnson:2019ljv, Gralla:2019ceu}.  If
$\Omega_{\rm ph}$ is its orbital frequency and $\lambda$ its radial
Lyapunov exponent, then
\begin{equation}
 \frac{\lambda}{\Omega_{\rm ph}}
 =\frac{\sqrt{3}(r_{\rm ph}-M)}{2\sqrt{Mr_{\rm ph}}}
 =\kappa+O(\kappa^2).
 \label{eq:lambdaOmega}
\end{equation}
The near extremal expansion further gives\begin{equation}\frac{\lambda}{\Omega_{\rm ph}}=\kappa-\frac{2\sqrt{3}}{9}\kappa^2+O(\kappa^3).\end{equation}

Over $n$ circular-orbit periods,
\begin{equation}
 n\lambda T_{\rm ph}=2\pi X+O(\kappa),
 \qquad T_{\rm ph}=\frac{2\pi}{\Omega_{\rm ph}},
 \label{eq:accumulated-instability}
\end{equation}
with the remainder understood at fixed $X$.  Thus the return
crossover occurs when the winding number becomes comparable to the
inverse instability per revolution,
$n\lambda T_{\rm ph}=O(1)$.  This comparison uses the circular-orbit
Lyapunov scale; it does not identify $\lambda$ with an exponent
integrated along each scattering trajectory.  For fixed subextremal
spin it also reproduces the standard strong-deflection ratio
$\epsilon_{n+1}/\epsilon_n\sim e^{-\lambda T_{\rm ph}}$ near the
critical orbit.

\emph{Note added. While an earlier version of the present work was under editorial consideration at Physical Review Letters, the preprint by Chowdhury \cite{Chowdhury:2026zbz} appeared, independently investigating the algebraic-to-exponential crossover of equatorial photon returns in near extremal Kerr spacetime. Beyond this radial crossover, the present work establishes a distinct crossover in the polar Jacobi phase and identifies the parametrically distinct scales governing photon returns and polar focusing.}

\section{Discussion}
\label{sec:discussion}
The near extremal limit reorganizes two different pieces of the lensing
geometry.  The radial tuning of equatorial returns crosses from the
extremal inverse-winding law to subextremal exponential accumulation at
$n\sim\kappa^{-1}$.  The polar Jacobi map of neighboring rays exhibits
a second crossover: its conjugate points pass from geometric spacing
in unwrapped winding to asymptotically constant spacing
$\Delta\mathcal N\simeq(2\kappa)^{-1}$, while the associated focusing
order grows as
$m_{\rm cross}^{(\theta)}\sim(2/\pi)\ln\kappa^{-1}$.  The two scales are
therefore parametrically different.  Near extremal Kerr can look
extremal in the radial return hierarchy over a very broad range of
windings even while the polar phase has accumulated many oscillations.

Both results have a controlled domain of validity.  The leading
scaling functions apply as $\kappa\to0$ at fixed positive finite
$X=n\kappa$, with a return radius $r_s/M>1$ held fixed.  In the radial
problem the exterior matching produces the logarithmically enhanced
correction in Eq.~\eqref{eq:intro-scaling}; in the polar problem it
produces the finite source-dependent constant $C_s$.  If instead
$r_s-M=O(\kappa M)$, the scaled endpoint remains in the near-horizon
region and both matching problems change.  Likewise, the asymptotic
remainders quoted here are not uniform at the endpoints
$X\to0$ and $X\to\infty$.

The focusing condition in Eq.~\eqref{eq:focusing-crossover} is a
linear polar conjugate-point condition about the equatorial family,
not by itself a full two-dimensional Kerr caustic or a prediction for
image brightness.  Generic quasi-equatorial strong-field lensing and
caustics are well established \cite{Bozza:2002af}, and the exactly
extremal strong-deflection problem has recently been extended to
magnifications and caustic positions \cite{Feleppa:2026emm}.
The contribution here is the controlled near extremal interpolation of
the radial return and polar focusing sectors in the same double-scaling
limit.  Embedding Eq.~\eqref{eq:focusing-crossover} into a complete
source-observer lens map is the next step required for flux or
finite-source predictions.

The mechanism suggests a broader asymptotic principle.  When a family
of unstable photon orbits approaches a degenerate horizon, different
components of the geodesic deviation problem need not share the same
critical scale.  In Kerr, the azimuthal return hierarchy is governed
by $n\kappa$, whereas polar focusing introduces the exponentially
rescaled variable $Z_m=\kappa e^{(m\pi-C_s)/2}$.  The near extremal
geometry therefore supports coupled but inequivalent winding and
focusing crossovers.  Determining whether analogous pairs of scaling
variables arise for generic nonequatorial Kerr rays or other
degenerate rotating horizons is a natural test of the universality of
this structure.

%\begin{acknowledgments}
%The author thanks  
%\end{acknowledgments}

\end{document}